\documentclass[manuscript]{aastex}

\newcommand\ha{H$\alpha$}
\newcommand\ca{\mbox{Ca\,{\sc ii}}}

\newcommand\si{\mbox{Si\,{\sc i}}}
\newcommand{\kms}{~km~s$^{-1}$}

\usepackage[normalem]{ulem}      
\usepackage{color}          

\shorttitle{jet above the light bridge}
\shortauthors{Song et al.}

\begin{document}


\title{Chromospheric Plasma Ejections in a Light Bridge of a Sunspot}


\author{Donguk Song\altaffilmark{1}, Jongchul Chae\altaffilmark{1}, Vasyl Yurchyshyn\altaffilmark{3}, Eun-Kyung Lim\altaffilmark{2}, Kyung-Suk Cho\altaffilmark{2}, Heesu Yang\altaffilmark{1}, Kyuhyoun Cho\altaffilmark{1}, and Hannah Kwak\altaffilmark{1}}

\email{dusong@astro.snu.ac.kr}


\altaffiltext{1}{Astronomy Program, Department of Physics and Astronomy, Seoul National University, 1 Gwanak-ro, Gwanak-gu, Seoul 08826, Korea}

\altaffiltext{2}{Korea Astronomy and Space Science Institute 776, Daedeokdae-ro, Yuseong-gu, Daejeon 34055, Korea}

\altaffiltext{3}{Big Bear Solar Observatory, New Jersey Institute of Technology, 40386 North Shore Lane, Big Bear City, CA 92314-9672, USA}

\begin{abstract}
It is well-known that light bridges inside a sunspot produce small-scale plasma ejections and transient brightenings in the chromosphere, but the nature and origin of such phenomena are still unclear. Utilizing the high-spatial and high-temporal resolution spectral data taken with the Fast Imaging Solar Spectrograph and the TiO 7057 \AA\ broadband filter images installed at the 1.6 meter New Solar Telescope of Big Bear Solar Observatory, we report arcsecond-scale chromospheric plasma ejections (1.\arcsec7) inside a light bridge. Interestingly, the ejections are found to be a manifestation of upwardly propagating shock waves as evidenced by the sawtooth patterns seen in the temporal-spectral plots of the \ca~8542~\AA\ and \ha\ intensities. We also found a fine-scale photospheric pattern (1\arcsec) diverging  with a speed of about 2 \kms\ two minutes before the plasma ejections, which seems to be a manifestation of magnetic flux emergence. As a response to the plasma ejections, the corona displayed small-scale transient brightenings. Based on our findings, we suggest that the shock waves can be excited by the local disturbance caused by magnetic reconnection between the emerging flux inside the light bridge and the adjacent umbral magnetic field. The disturbance generates slow-mode waves, which soon develop into shock waves, and manifest themselves as the arcsecond-scale plasma ejections. It also appears that the dissipation of mechanical energy in the shock waves can heat the local corona.



\end{abstract}



\section{INTRODUCTION}

Light bridges (LBs) are filamentary bright structures dividing a sunspot umbra into two umbral regions with the same magnetic polarity \citep{muller79}. They comprise small-scale successive convective cells and an elongated dark lane running parallel to the central axis. A LB has a special status in that convective motions are not fully inhibited by magnetic fields \citep{hirzberger02,rimmele08}. Plasma below it can penetrate into the vertical magnetic fields of umbra \citep{schussler06}, and the magnetic fields above it form a magnetic canopy structure \citep{jurcak06}. Such a complex topology of the magnetic fields and magnetoconvection in the LB are believed to be responsible for a variety of dynamic phenomena in the chromosphere, but the nature and origin of the chromospheric activities in the LBs still remain unresolved.

The most well-known chromospheric activities in the LB are plasma ejections seen as jet-like features in the \ha\ \citep {roy73,asai01} and \ca~H \citep{shimizu09,shimizu11,louis14} images. The plasma ejections intermittently recur with a short lifetime more than one day \citep{shimizu09}, and their length are usually shorter than 1000~km \citep{louis14}. 
The plasma ejections are generally observed at one side of the LB, and the ejected plasma moves along the magnetic field lines above the LB \citep{asai01,louis14}. Meanwhile, \citet{asai01} reported loop-like bright structures in the {\it TRACE} 171~\AA\ image that traced the edge of plasma ejections seen in the \ha\ image. They suggested that the loop-like structures indicate the emergence of the bipolar magnetic flux in the LB. \citet{liu12} found a coronal jet ejected from a LB in the EUV 171 \AA\ image. They inferred that the coronal jet originated from magnetic reconnection between the LB and the sunspot umbra.

Magnetic reconnection in the low chromosphere and the upper photosphere of a LB is thought to be responsible for the chromospheric plasma ejections above the LB. \citet{bharti07} reported an opposite magnetic polarity inside a LB and suggested that the low-altitude magnetic reconnection above the LB may help dense plasma to be injected into the chromosphere. \citet{shimizu09} found a strong electric current sheet formed along a LB. They proposed that highly twisted flux tubes trapped below the magnetic canopy are favorable for the magnetic reconnections inside the LB. Meanwhile, \citet{louis15} found a small-scale, flat $\Omega$-loop emerging from a granular LB that was associated with the chromospheric emission detected at one side of the $\Omega$-loop.

High-resolution observations of LBs from the {\it Hinode} \citep{tsuneta08} and Interface Region Imaging Spectrograph \citep[IRIS;][]{pontieu14} revealed enhanced brightenings at the edge of jets seen in a LB and coordinated behaviors with the adjacent jets \citep{bharti15}. \citet{bharti15} speculated that waves leaking from the LB produce these kinds of coordinated behaviors between jets above the LB and adjacent jets. The excitation of waves in the low chromosphere above the LB can be regarded as one of the possible scenarios for the generation of plasma ejections. This has been reported by the previous studies of dynamic jet-like features seen at the region outside sunspots that slow-mode shock waves excited in the low chromosphere can play an important role for the evolution of dynamic jet-like features, such as dynamic fibrils \citep{hansteen06,dep07,langangen08}, superpenumbral fibrils \citep{chae14,chae15b}, and surges/jets \citep{morton12,yang14}. In this regard, we expect that shock waves genereated in the low chromosphere of a LB may appear as fine-scale chromospheric plasma ejections in the LB, but there have been no clear observational reports on the relation between shock wave and chromospheric plasma ejections above a LB until now.

Here, we report that upwardly propagating shock waves excited in the low chromosphere appear as arcsecond-scale chromospheric plasma ejections above a LB. For this study, we combine the high resolution spectral data taken with the Fast Imaging Solar Spectrograph \citep[FISS;][]{chae13} installed at the 1.6 meter New Solar Telescope \citep[NST;][]{goode10} of Big Bear Solar Observatory (BBSO) and the extreme-ultraviolet (EUV) data taken by the Atmospheric Image Assembly \citep[AIA;][]{lemen12} on board the {\it Solar Dynamics Observatory} \citep[SDO;][]{pesnell12}. We investigate in detail the spectral characteristics of the chromospheric plasma ejections detected above the LB and an unique photospheric flow pattern associated with the chromospheric plasma ejections.
Our paper is organized as follows. In Section 2, we describe observations and data analysis of the LB. Our observational findings are presented in Section 3. In Section 3.1, shock wave phenomena of the chromospheric plasma ejections are described. The description of the photospheric flow pattern associated plasma ejections is also presented in Section 3.2. In Section 3.3, the response of the corona to the plasma ejections is described. Finally, we summarize and discuss the physical implications of the chromospheric plasma ejections above the LB in Section 4.

\section{OBSERVATIONS AND DATA ANALYSIS}

\begin{figure}
\plotone{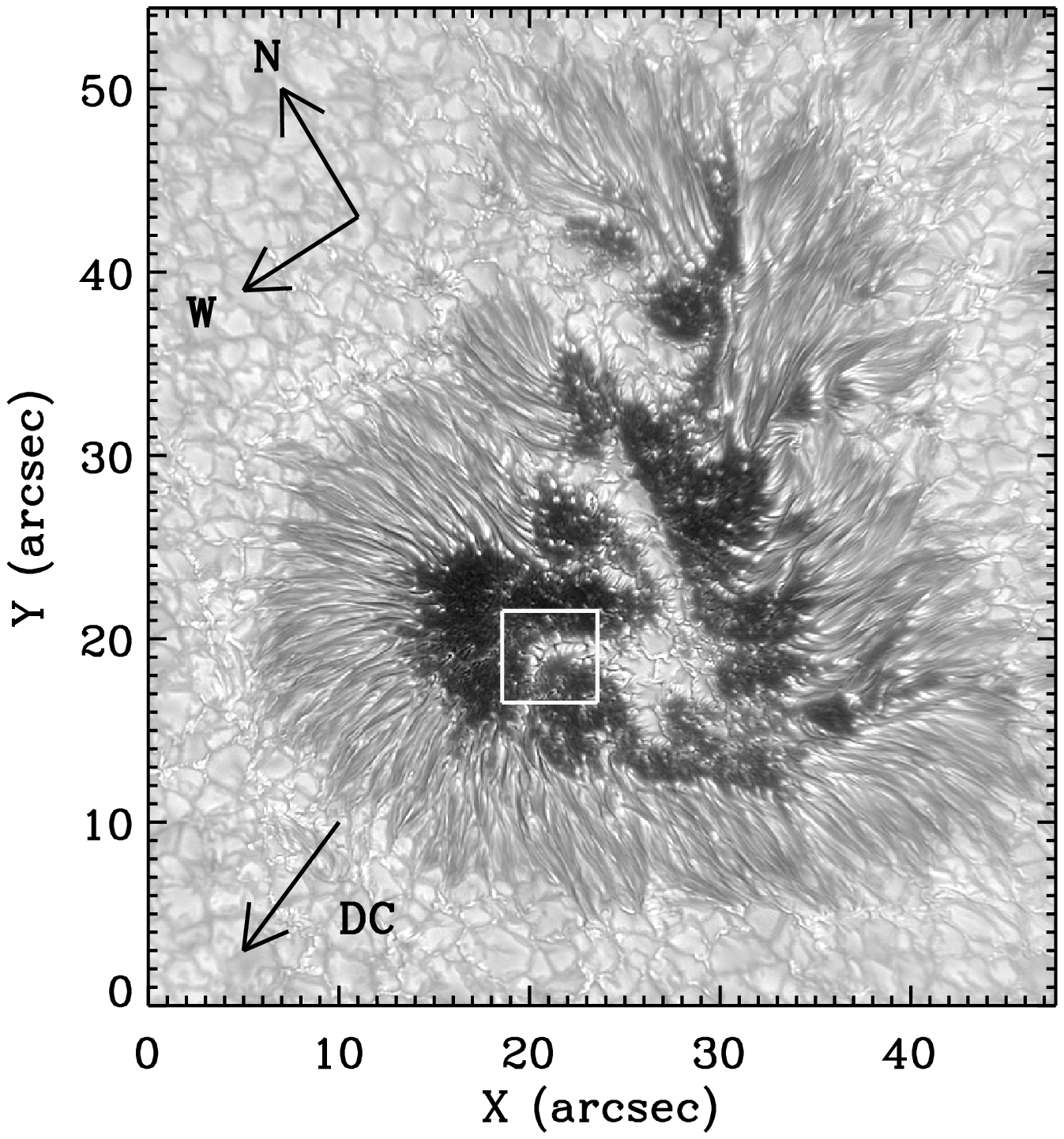}
\caption{Photospheric NST/TiO 7057 \AA\ image (bandpass: 10 \AA\ $\&$ pixel size: 0.034\arcsec) of the leading sunspot of NOAA AR 12082 (N17, E25) observed on June 6, 2014 at 20:52:18 UT. The rectangle represents the field-of-view of the region of our interest seen in Figure 2. DC indicates the direction of solar disk center.}\label{fig1}
\end{figure}

On 2014 June 6, we observed a leading sunspot (Figure~\ref{fig1}) in NOAA Active Region 12082 (N17, E25) with the FISS and the TiO~7057~\AA\ broadband filter \citep{cao10} installed at the 1.6 meter NST of BBSO. The FISS is a high dispersion Ech{\'e}lle spectrograph that can simultaneously record the \ha\ and \ca\ 8542 \AA\ lines based on the fast scanning mode. The spectrograms of the \ha\ and \ca\ lines respectively cover the wavelength ranges from -5~\AA\ to 5~\AA\ and from -8~\AA\ to 5~\AA\, with the spectral dispersion being 0.019~\AA\ and 0.026~\AA. The slit size is 0.$''$16 wide and 40$''$ long. One scan of the raster image  comprises 160 steps and covers a 25$''$$\times$40$''$ field of view (FOV). The exposure time was 30~ms. It took about 23~s for completion of one scan. The basic data processing including flat-fielding, distortion correction, compression, and noise reduction was done in the way described by \citet{chae13}. The line-of-sight (LOS) Dopplergrams of the photosphere and the chromosphere were computed from the lambdameter method \citep{deubner96} at multi-wavelength levels of the \ha\ and \si~8536~\AA\ lines. 

Figure \ref{fig1} shows the photospheric image of the sunspot that we observed at 20:52:18 UT on June 6, 2014. It was taken by the TiO 7057 \AA\ broadband filter with the aid of 308 sub-aperture adaptive optics \citep{shumko14}. The speckle reconstruction of the TiO image was done by using the Kiepenheuer-Institut Speckle Interferometry Package code reported by \citet{woger08}. We obtained the photospheric images every 15~s from 20:04:50 UT to 21:04:32 UT. The sunspot was rapidly growing during this observational time. It had the varied shapes of LBs dividing a sunspot umbra into several umbral regions. A white rectangle in Figure~\ref{fig1} shows the region of our interest, covering 5\arcsec$\times$5\arcsec. The region includes a strong LB embedded in a dark umbra. The LB has negative-polarity magnetic fields with the field strength below 1200~G. The surrounding umbra has magnetic field of the same polarity stronger than 1400~G. From the TiO image, we find a dark lane running parallel to the central axis of the LB and a number of small-scale successive convective cells.

The coronal images were acquired with the AIA on board the SDO. The SDO/AIA provides high resolution full-disk data of the Sun in 7 extreme-ultraviolet (EUV; $\lambda$94 \AA, 131 \AA, 171 \AA, 193 \AA, 211 \AA, 304 \AA, 335 \AA) and 3 ultraviolet (UV; $\lambda$1600 \AA, 1700 \AA, and 4500 \AA) channels with the pixel scale of 0.$''$6 at a time cadence of 12~s (24~s in UV channels). To investigate temperature variations of EUV-emitting plasma in the corona, we determined the filter ratios of the 171 \AA, 193 \AA, and 211 \AA\ passbands, which have been often used in a determination factor of temperatures in the corona \citep{chae02}. In particular, the variations of the filter ratios (193 \AA/171 \AA\ and 211 \AA/171 \AA) are very sensitive to the temperature variations of 1$-$2 MK.

The FISS data were aligned with the SDO/Helioseismic and Magnetic Imager \citep[HMI;][]{schou12} data by using cross-correlation between the SDO/HMI intensity image and the FISS/\ca~-4~\AA\ image, and then the co-alignment between the SDO/HMI and the SDO/AIA data were carried out by using the IDL (Interactive Data Language) programs of the hmi$\_$prep.pro and the aia$\_$prep.pro.

\section{Results}

\subsection{Chromospheric Plasma Ejections}

\begin{figure}
\plotone{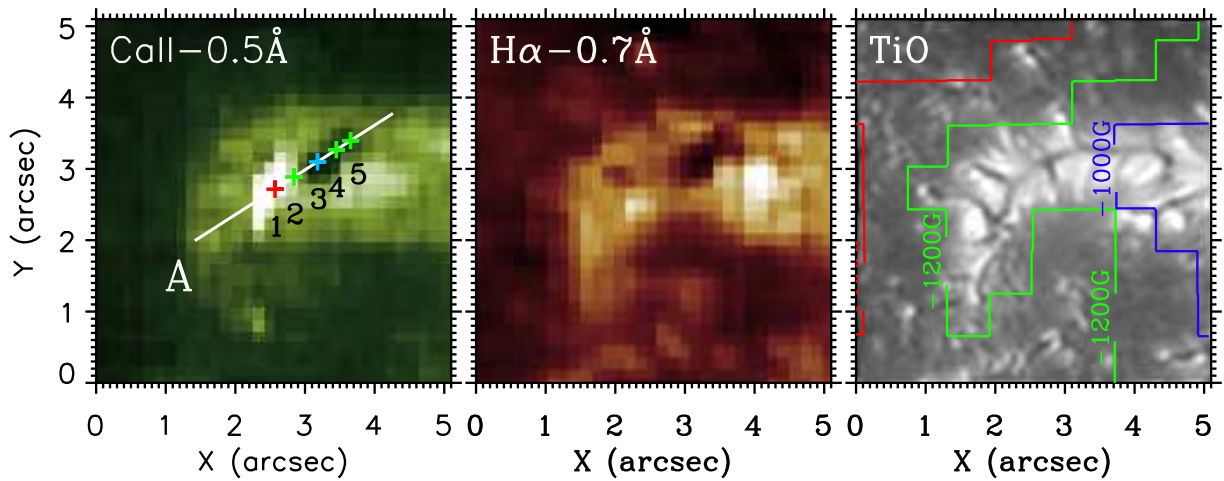}
\caption{Chromospheric and photospheric images of the region of our interest taken with the FISS and the TiO broadband filter at 20:52:24 UT. The FOV is 5$''$$\times$5$''$ corresponding to the white rectangle seen in Figure \ref{fig1}. The contours seen in the TiO image presents the magnetic field strength that is scaled in the range from -1000 G (blue line) to -1400 G (red line). The field strength was determined from the SDO/HMI magnetogram. (An animation of this figure is available.) 
}\label{fig2}
\end{figure}

A chromospheric plasma ejection in an arcsecond-scale is visible above a LB in the \ca~-0.5 \AA\ and \ha~-0.7 \AA\ images taken by the FISS (Figure~\ref{fig2}). The maximum length of the plasma ejection is about 1250~km (1.\arcsec7), and the width is about 290~km (0.\arcsec4). In the \ca~-0.5~\AA\ image, the chromospheric plasma ejection divides into two distinct parts; a bright patch and a dark patch. On the other hand, the bright patch is not conspicuous in the \ha~-0.7~\AA\ image. This may be attributed to the difference in temperature sensitivity between the \ca\ line and the \ha\ line \citep{cauzzi09}. \ca\ intensity is more sensitive to the temperature of chromospheric structures than \ha\ intensity, while \ha\ intensity is more subject to the light-scattering of the chromospheric features \citep{chae13a}. 

\begin{figure}
\includegraphics[width=0.95\textwidth,clip=]{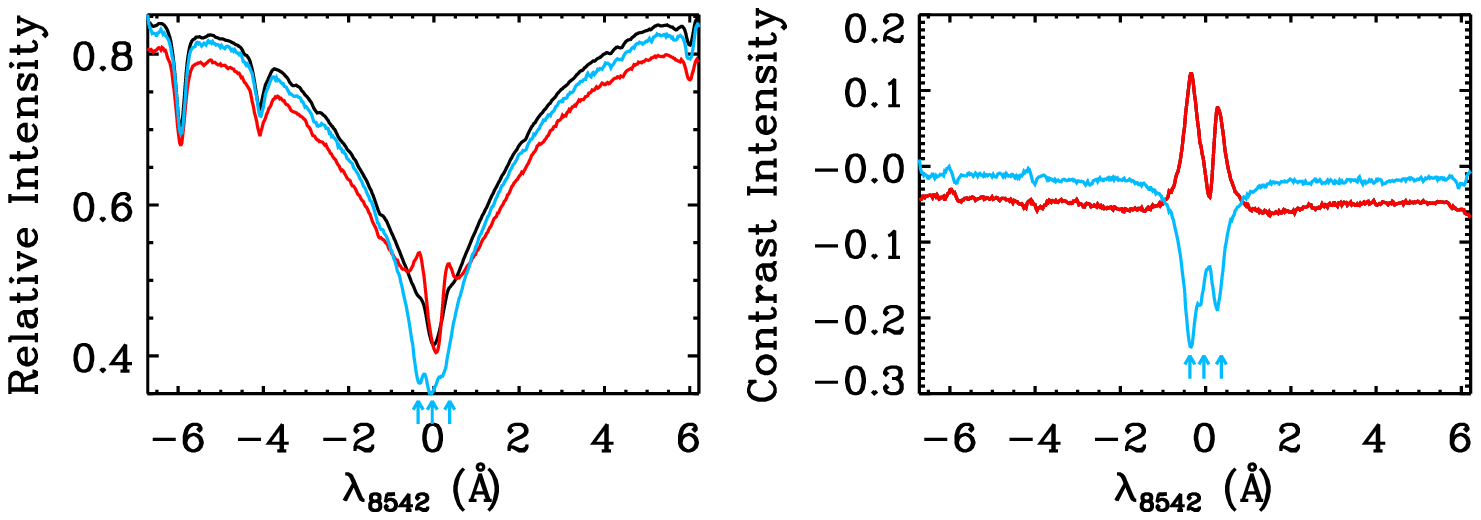}
\includegraphics[width=0.95\textwidth,clip=]{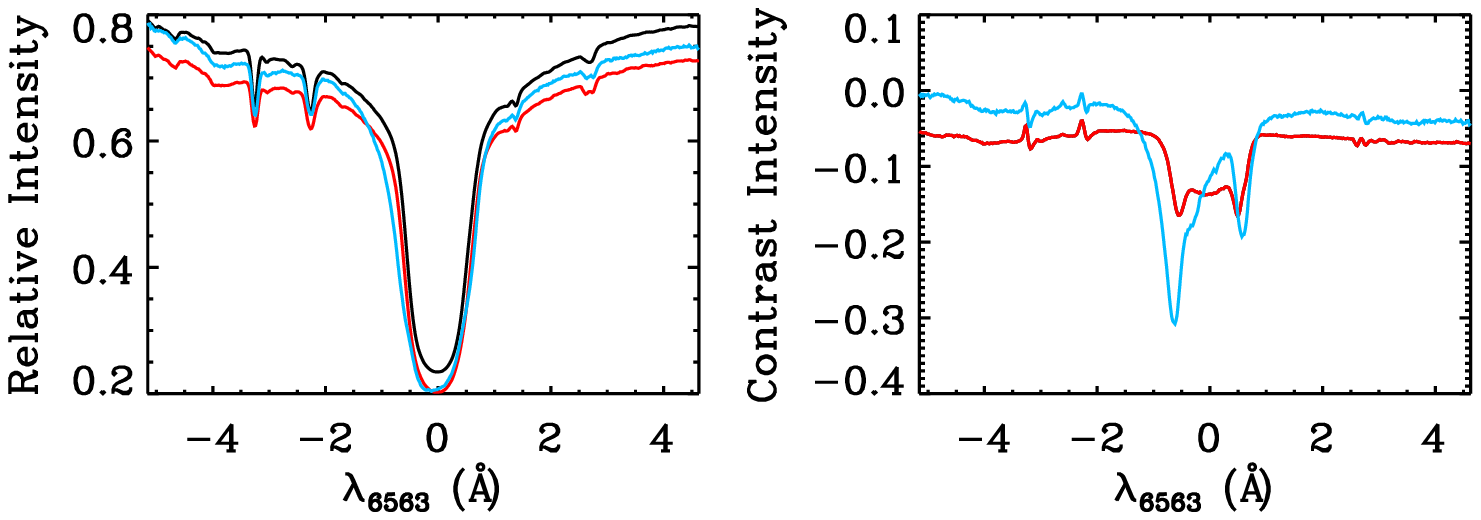}
\caption{\ca\ (top panels) and \ha\ (bottom panels) spectral (left panels) and contrast (right panels) profiles of the chromospheric plasma ejection at 20:52:24 UT. The spectral and contrast profiles are constructed at the bright (red) and dark (cyan) patches in the plasma ejection, which are denoted by the red (point of 1) and cyan (point of 3) plus symbols in Figure~\ref{fig2}. The black solid lines in the left panels present average profiles constructed in the region inside the LB.}\label{fig3}
\end{figure}

The bright patch and the dark patch of the plasma ejection have different spectral characteristics in both the lines (Figure~\ref{fig3}). First, the \ca\ spectral profile of the bright patch shows strong emission  in the blue and red wings at the wavelength ranges [-0.4~\AA\ and -0.3~\AA] and [+0.3~\AA\ and +0.4~\AA], respectively. Meanwhile, we find a weak absorption feature in the \ca\ line core. This shape of the \ca\ spectral profile -- a broad Gaussian emission at the \ca\ line wings and a central absorption at the \ca\ line core -- corresponds to the shape of the \ca\ spectral profile at a transient brightening above a LB reported by \citet{louis15}. In addition, it is similar to the well-known \ca\ spectral profiles of the Ellerman bombs \citep{ellerman17} and the penumbral microjets \citep{katsukawa07,vissers15} located outside sunspot umbrae.

We also find from the \ca\ spectral profile that the blue wing has stronger emission than the red wing. This asymmetric emission is more pronounced in the \ca\ contrast profile. The asymmetric wing emission generally reflects mass motion at the event formation height, and so we infer that upward plasma motion was predominant in the low chromosphere. In the \ha\ spectral profile, however, such a strong emission feature is not noticeable. This is different from the case of an Ellerman bomb. Second, the \ca\ spectral profile of the dark patch in the plasma ejections, presents several absorption cores (cyan arrows in Figure~\ref{fig3}). These absorption cores are especially well-identified in the contrast profiles of the \ca\ and \ha\ lines. Our data suggest that the dark patch of the plasma ejection is composed of plasmas that have several velocities.

\begin{figure}
\epsscale{.70}
\plotone{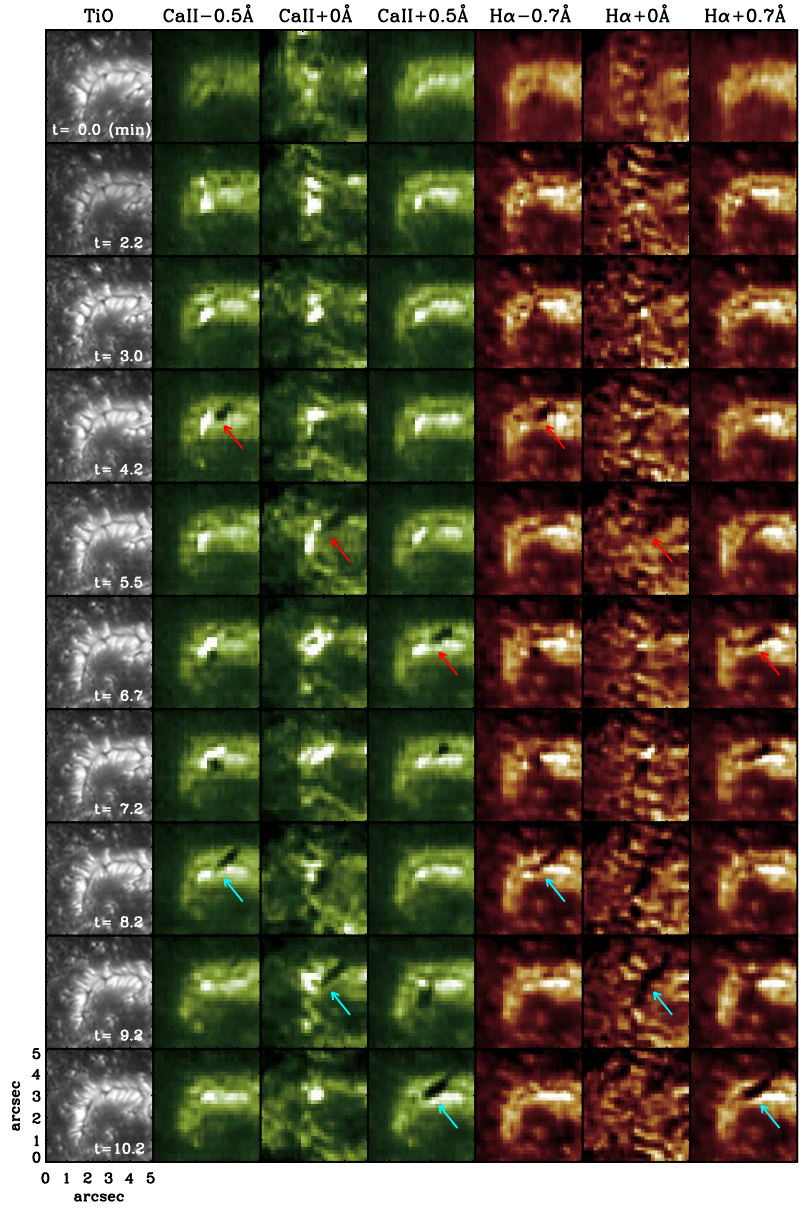}
\caption{Temporal evolutions of  the photospheric TiO images (Column 1) and plasma ejections obtained with the different wavelengths of the \ca\ (Columns 2-4) and \ha\  (Columns 5-7) in the LB from 20:48:04 UT to 20:58:18 UT. Red arrows present the first event of the plasma ejection, and cyan arrows indicate the second event.}\label{fig4}
\end{figure}

Figure~\ref{fig4} shows the temporal variations of the chromospheric plasma ejections seen in the TiO, \ca\, and \ha\ images from 20:48:04 UT to 20:58:18 UT. We find from the figure that the plasma ejections in the LB occurred twice within 10 minutes. Rows $4-6$ in Figure~\ref{fig4} present the first plasma ejection, and rows $7-10$ present the second plasma ejection. Each ejection lasts about 3 minutes. Moreover, we find that the dark patch in the plasma ejections are seen successively at the blue wings, the cores, and the red wings (see, red and cyan arrows in Figure~\ref{fig4}), which shows that the plasma motion switched from upward to downward directions.

\begin{figure}
\includegraphics[width=0.9\textwidth,clip=]{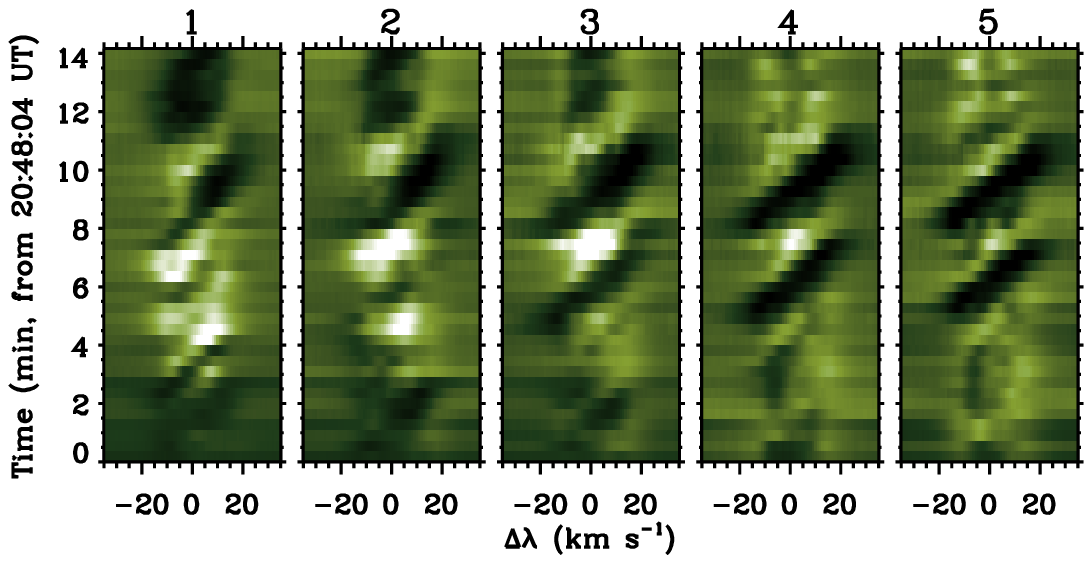}
\includegraphics[width=0.9\textwidth,clip=]{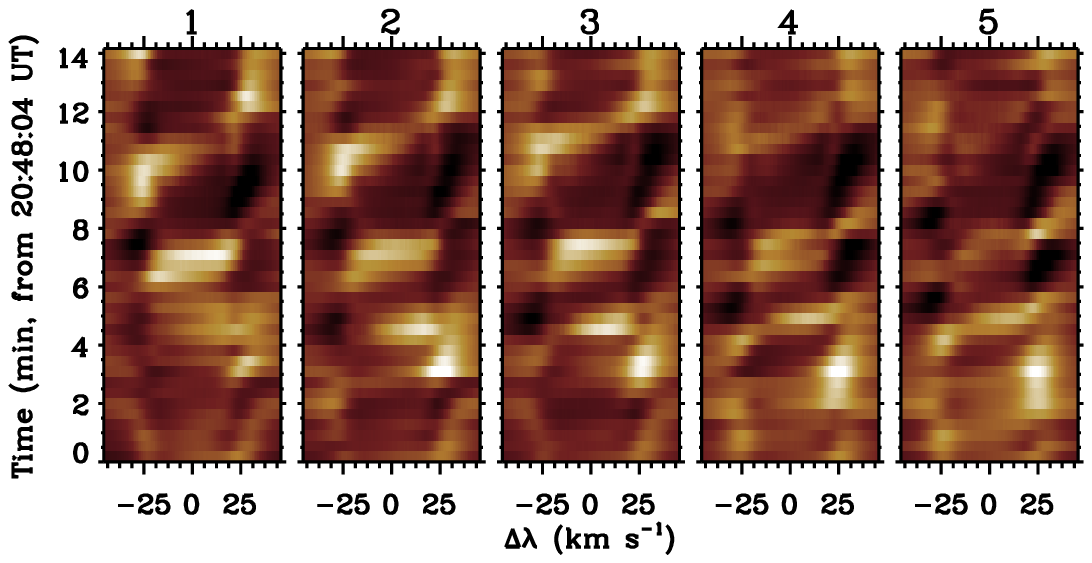}
\caption{Wavelength-time plots ($\lambda-t$ plots) of the \ca\ (top) and \ha\ (bottom) contrast profiles constructed at five different position heights of the plasma ejections. Each panel corresponds to the corresponding number given in the \ca\ -0.5 \AA\ image of Figure~\ref{fig2}. The negative values indicate the upward motion (blue-shifted direction) of the ejected plasma, while the positive values represent the downward motion (red-shifted direction). }\label{fig5}
\end{figure}

\begin{figure}
\epsscale{.80}
\plotone{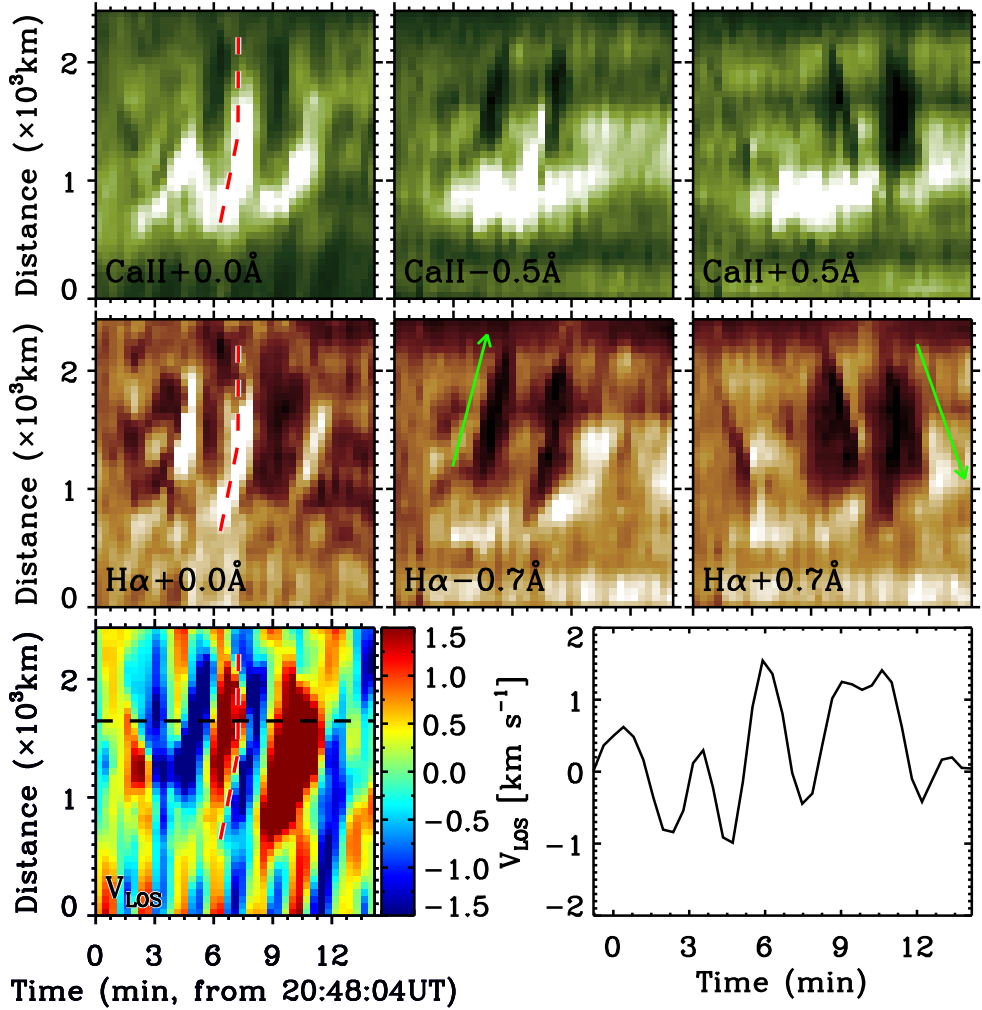}
\caption{Time-distance plots ($t-d$ plots) of the \ca\ (top) and \ha\ (middle) intensities and the LOS Dopper velocity (bottom) determined by applying the lambdameter method to the \ha\ line. These plots are constructed from the slit $'$A$'$ running parallel to the plasma ejection denoted in Figure~\ref{fig2}. The right pannel of the bottom presents of the temporal variation of LOS velocity constructed from the black dashed line denoted in $t-d$ plot of the velocity. The distance of 0 km in the $t-d$ plots corresponds the lower-left of the slit in Figure~\ref{fig2}.}\label{fig6}
\end{figure}

Figures~\ref{fig5} and Figure~\ref{fig6} present the spatio-temporal patterns of the plasma ejections in the wavelength-time plots ($\lambda-t$ plots) and the time-distance plots ($t-d$ plots) of the spectral data. The observed patterns  are  quite like what are expected in shock waves propagating upwards in several aspects. 
First,  the $\lambda-t$ plots indicate the temporal pattern of downflow, emission, and upflow. This kind of pattern was repeated at least three times during our observationas, which constitute a periodic pattern of about three-minute period. In particular, the two recurrent plasma ejections correspond to the upflow phases in two strongest patterns.  
Second, each pattern includes a velocity jump from fast downflow (large redshift) to fast upflow (large blueshift). It especially appears as a sawtooth or a N-shape pattern in the $\lambda-t$ plots. Third,  the velocity jump is accompanied by  an emission in the line cores.  The emission core reflects the increase of source function in the line, and may be considered as an evidence for strong compression and  heating. Finally,  the position of each  emission moves in the plane of sky with a speed of 10.5 \kms\ or higher. In particular, the propagation of each emission is observed at the boundary between the large redshift and the large blueshift, which is regarded as shock frount (see, Figure~\ref{fig6}). 

Figure~\ref{fig5} confirms that each plasma ejection starts with a sudden appearance of upward motion,  deceleration,  gradual switch to downward motion, and then downward acceleration.  The peak speed of upflow  is  estimated at about 26 \kms\  in the \ha\ data and 20 \kms\ in the \ca\ data, which turns into the downflow of the same speed in about three minutes. Therefore the resulting deceleration is about 290~m~s$^{-2}$ and 220~m~s$^{-2}$, respectively. These values  are close to solar gravitational acceleration 274~m~s$^{-2}$, but  the proximity may be a coincidence,  because the measured deceleration is  determined by the peak speed for the given period of three minutes.  The positive correlation between peak speed and deceleration was previously reported in dynamic fibrils and was regarded as evidence for shock waves based on the comparison with numerical simulations  \citep{hansteen06,dep07}.

\subsection{Pattern of Photospheric Diverging Motion}

We find an unique photospheric pattern of diverging motion (1\arcsec) in the surface of the LB before the occurrence of the chromospheric plasma ejections. This is well identified in the time sequence of TiO images from 20:50:18 UT (t=2.2 min) to 20:52:18 UT (t=4.2 min) shown in Figure~\ref{fig8}. In the figure, the photoshperic pattern first appeared near the dark lane of the LB, and then expanded across the LB. The expanding direction of the pattern is almost the same as the direction of the ejected plasma in the chromosphere. Two small bright points with the diameter of 90~km are visible at the terminations of the photospheric pattern. They are comparable to quiet Sun bright points in the intergranular lanes in size. The chromospheric plasma ejections of our interest occurred at the one side of the photospheric pattern when the edge of the pattern reached the boundary of the sunspot umbra (see, Figure~\ref{fig2}). Our data suggest that the chromospheric plasma ejections are closely related with the diverging motion in the photosphere.

The photospheric pattern of diverging motion mentioned above is well identified in the time-distance plot ($t-d$ plot) of the TiO intensity in Figure~\ref{fig9}. The tilted bright strands indicate the trajectories of two bright points located at the termination positions of the diverging pattern. Its horizontal speed is about 2~\kms, which is faster than the typical speed of horizontal flows in a LB \citep{louis14}, but is slower than the speed of the expanding granular cells seen in emerging flux regions outside sunspots \citep{yang13,kim15}. In the $t-d$ maps of the \ca~$\pm$0.5~\AA\ and core intensities of Figure~\ref{fig9}, we can see successive enhanced brightenings next to the one side of the photospheric pattern. These intense brightenings are the lower part of plasma ejections. We find from Figure~\ref{fig9} that the brightenings in the \ca~-0.5~\AA\ image are more intense than that of the \ca~+0.5~\AA\ image. Since the intense brightenings in the line wing images present the plasma motion and temperature in the low chromosphere, we suggest that the plasma in the low chromosphere mostly have upward motion. We also find that the one side of the flow pattern corresponds to the strong redshift elongated feature in the plot of the photospheric Doppler velocity determined from the lambdameter method in \si\ line (Figure~\ref{fig9}). The redshift feature indicates downflows in the photosphere. The peak speed of the downflows is about 0.72~\kms. This value comparable to the the speed of redshift patch reported by \citet{shimizu09} as a photospheric counterpart of the bi-directional jet occurring in a LB.

\subsection{Response of the Corona to the Plasma Ejections}

As a response to the chromospheric plasma ejections in the LB, the corona displays small-scale transiently enhanced brightenings. In particular, these brightenings are seen in all the SDO/AIA images (Figure~\ref{fig10}). It indicates that the brighteings have a large range of temperatures from 2 MK to 20 MK. The temporal variation of the SDO/AIA 171~\AA\ intensity from 20:48:04 UT to 20:58:18 UT (Figure~\ref{fig10}) presents that the brightening appeared at least twice within 10 minutes. Each event time corresponds to the occurrence time of each plasma ejection in the chromosphere. Our data suggest that the coronal brightenings are intimately related with the occurrence of the plasma ejections in the chromosphere.

\section{Discussion}

We have reported observational results on  arcsecond-scale plasma ejections ($\sim$1.\arcsec7) in the chromosphere above a LB. Our findings are quite consistent what are expected for the upwardly propagating shock waves, suggesting that each plasma ejection may correspond to a flow event that starts with a sudden appearance of high speed upflow (following the shock front) and then decelerates until it changes to high speed downflow. This kind of shock wave  interpretation is quite new for plasma ejections in LBs, but was recently made quite often  in other jet-like features such as dynamic fibrils \citep{hansteen06,dep07,langangen08}, superpenumbral fibrils \citep{chae14},   small surges \citep{yang14}, and newly discovered umbral spikes \citep{yurchyshyn14}.

One may think that our observational results may be alternatively interpreted by the conventional picture of reconnection-driven ejections. According to this picture,  the line core emission  we observed might be considered as the evidence of direct heating from magnetic reconnection. The reconnection may also produce an impulsive acceleration which results in the sudden appearance of  high speed upflow, and its gradual deceleration by gravity.  The moving emission may represent a hot jet driven by high pressure  localized  in the reconnection site.  If this picture holds,  the pressure gradient force and the magnetic tension force may decrease with time. In observation, the emission feature should expand with distance as it moves  with  density and pressure  becoming lower and lower. This expectation is not compatible with  our finding that the moving emission feature continues to keep the condition of high compression implied by the velocity jump. Moreover, the reconnection picture has also a difficulty in explaining the  periodic pattern of emission and velocity.  One would attribute the periodic pattern to  the recurrent reconnection. But why and how does such reconnection takes place? Why at the three-minute period?

We have a way of explaining the periodic pattern using the shock wave picture.  We suppose  reconnection impulsively  takes place in the upper photosphere or in the low chromosphere.  This layer is highly stratified because of low temperature, and has an acoustic cutoff period of about three minutes.  In this kind of medium, 
 the impulsive disturbance of the medium produces a packet of waves with different frequencies. If the disturbance scale is large enough,  the power is concentrated in the waves with frequencies near the acoustic cutoff.  These waves are dispersive in that higher frequency waves have faster group speeds than lower frequency waves.  Thus a singe impulsive disturbance  produces a train of waves with the frequency slowly decreasing to the acoustic cutoff value  \citep{chae15a}. Each wave propagating upwards develops into a shock wave and manifests itself as a plasma ejection.  Therefore a single  impulsive reconnection event can produce a train of plasma ejections. The single impulsive reconnection event as mentioned above seems to be caused by the emergence of new magnetic flux into the pre-existing magnetic fields \citep[e.g.,][]{heyvaerts77}.   

\begin{figure}
\epsscale{.98}
\plotone{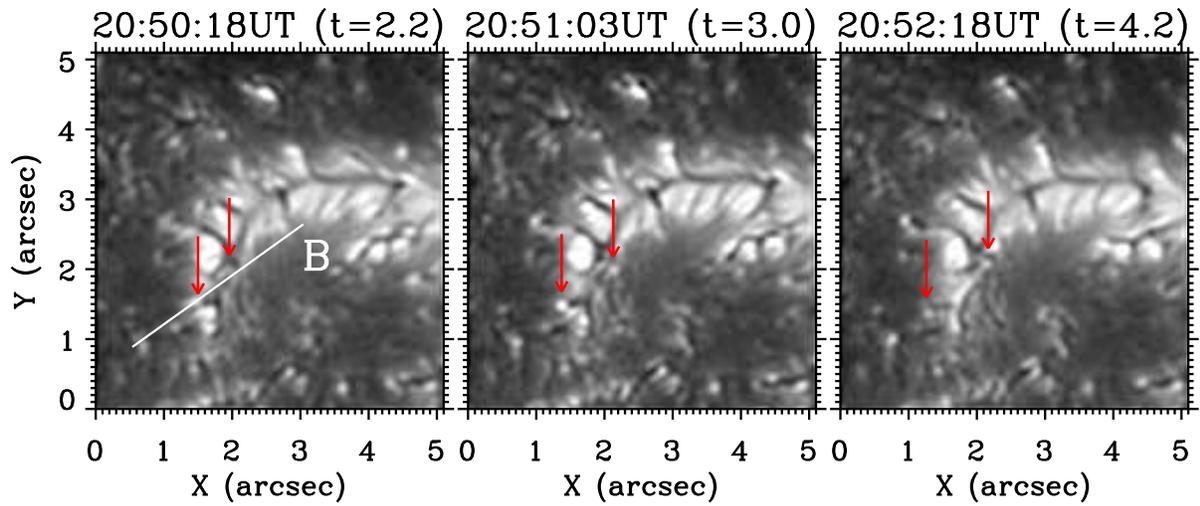}
\caption{TiO images of the region of our interest from 20:50:18 UT to 20:52:18 UT. Red arrows indicate the termination positions (small-scale bright points) of the photospheric diverging motion. The slit $'$B$'$ represents a position running parallel to the photospheric pattern of diverging motion.} \label{fig8}
\end{figure}

\begin{figure}
\epsscale{.98}
\plotone{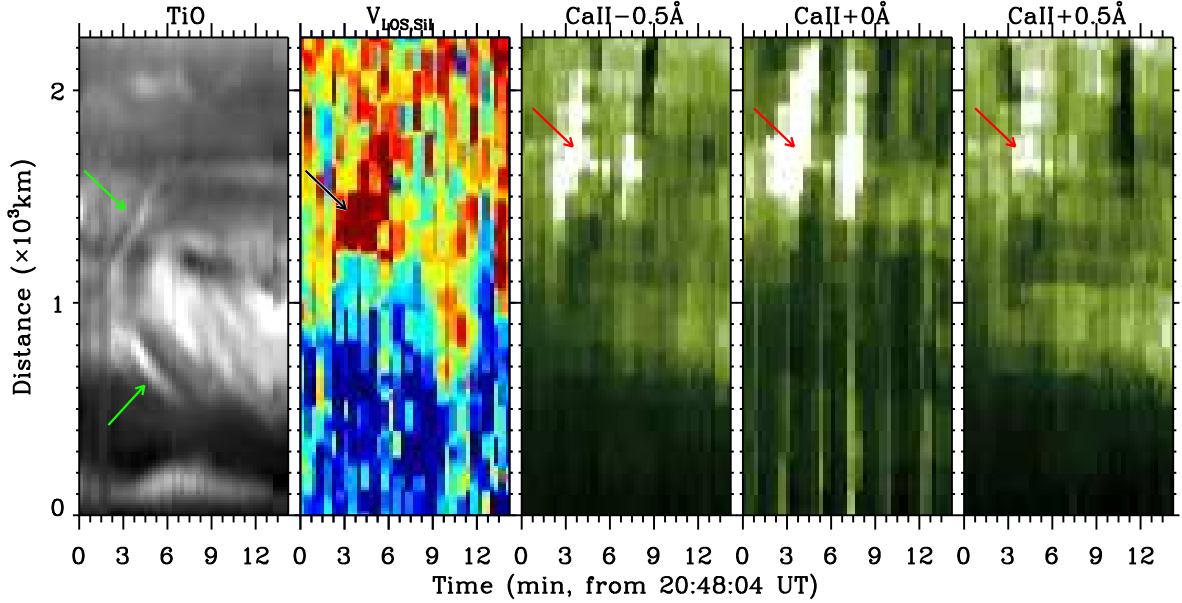}
\caption{Time-distance plots ($t-d$ plots) of the TiO intensity, the LOS Doppler velocity constructed from the \si\ line, and \ca\ -0.5~\AA\, core, +0.5~\AA\ intensities. They are constructed from the slit $'$B$'$ seen in Figure~\ref{fig8}. Green arrows present trajectories of the edges of the photospheric pattern of diverging motion. Black arrow indicates the downward motion seen along the one side of the photospheric diverging motion. Red arrows represent successive enhanced brightenings next to the one side of the pattern of the photospheric diverging motion. The distance of 0 km in the $t-d$ plots corresponds the lower-left of the slit in Figure~\ref{fig8}.} \label{fig9}
\end{figure}

\begin{figure}
\includegraphics[width=1.\textwidth,clip=]{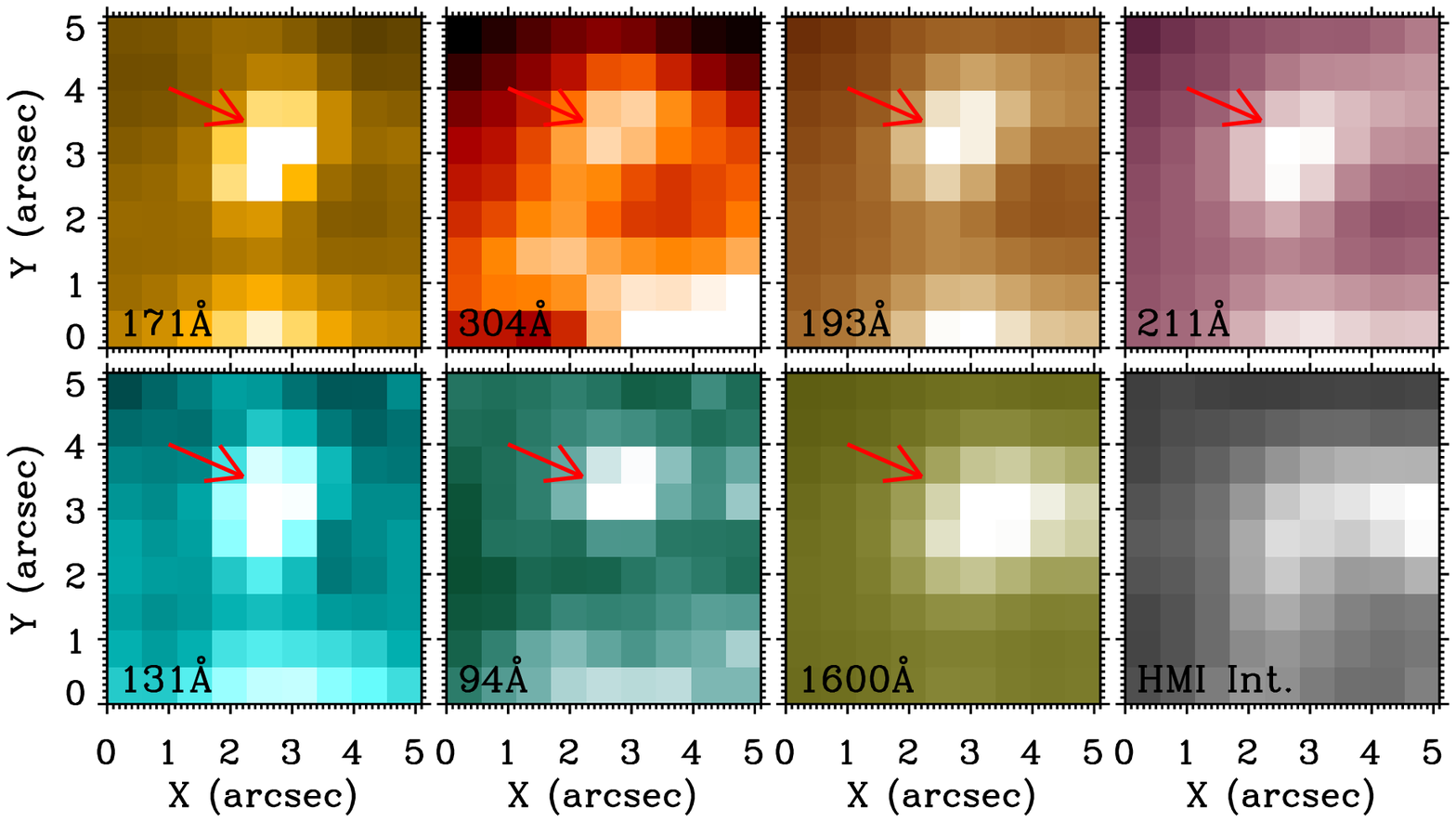}
\includegraphics[width=0.95\textwidth,clip=]{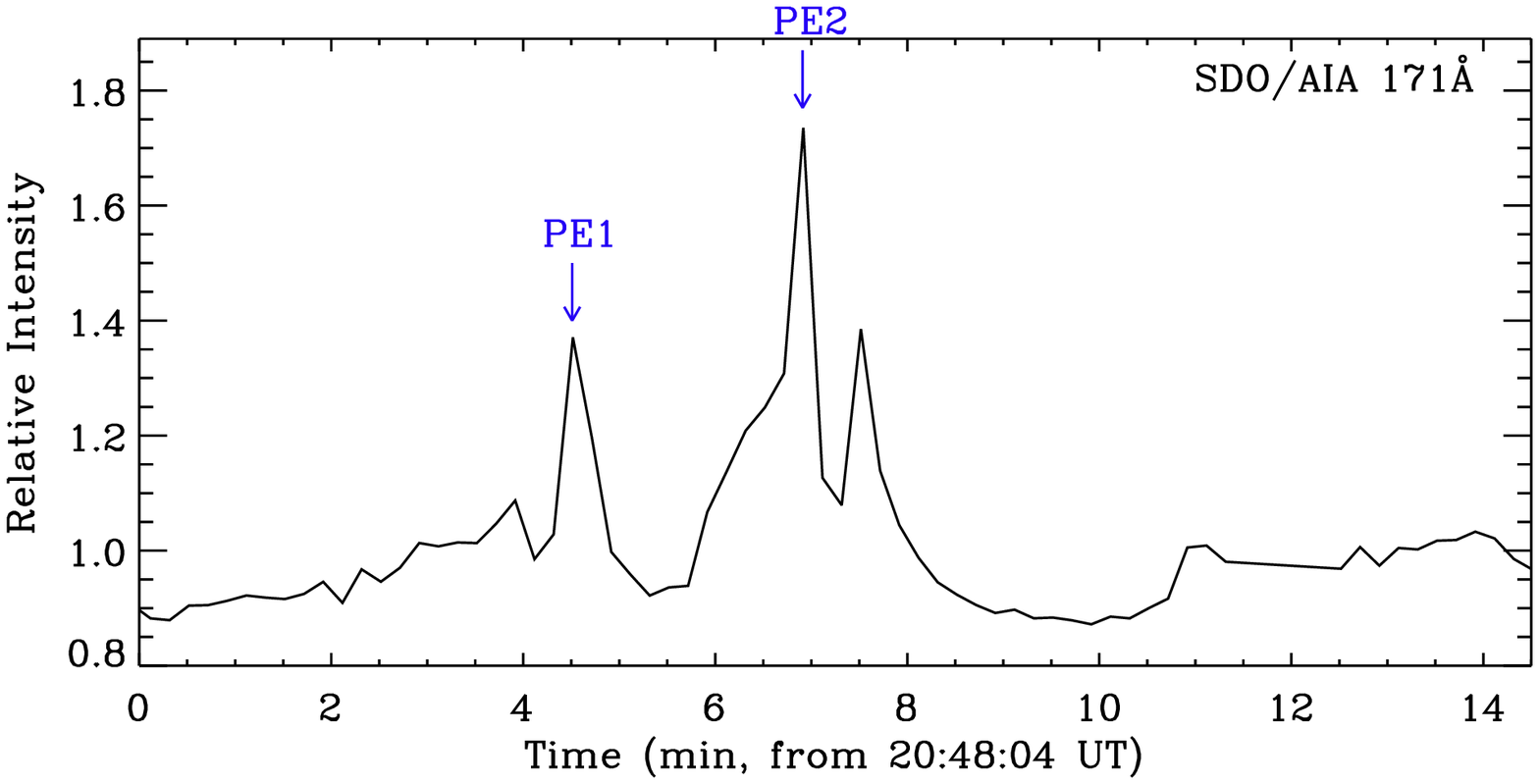}
\caption{Top panels: Coronal images of the region of our interest constructed from the SDO/AIA. Red arrows represent the small-scale brightenings that are associated with chromospheric plasma ejections. Bottom panel: Time variation of the SDO/AIA 171~\AA\ intensity. PE1 and PE2 indicate the first and second events of the plasma ejections.}\label{fig10}
\end{figure}

In our study, a manifestation of the emerging magnetic flux appears as a fine-scale photospheric pattern (1\arcsec) of the diverging motion in the sequence of TiO images (see, Figure~\ref{fig8}). In particular, the chromospheric plasma ejections we observed were only seen at the one side of the photospheric pattern, when the edge of the photospheric pattern reached at the boundary of the sunspot umbra. This is consistent with the previous studies on the photospheric counterpart of the magnetic flux emergence. \citet{otsuji07} reported that a horizontal expansion of granular cells in the photosphere is closely related with the emergence of the magnetic flux. 
\citet{lim11} investigated elongated granule-like features, so-called Moving Magnetic Features (MMF), stretching from the penumbral filaments. They mentioned that such photospheric features are related to the moving and emerging magnetic flux tubes. \citet{yang16} also reported several bi-directional expanding motions of granular-like structures in the photosphere, when the magnetic flux emerges below the Ellerman bombs. In this regard, it seems that the photospheric pattern of our interest represents a manifestation of the magnetic flux emergence arising in the LB, even though we do not have high-resolution magnetogram data that can directly confirm a pair of opposite polarities.

The coronal counterpart of the chromospheric plasma ejections displays small-scale transiently enhanced brightenings in the EUV images at all the coronal channels. During our observational time, the brightenings took place twice, and each brightening corresponds to 
each plasma ejection in the chromosphere. It is widely believed that the transient brightenings indicate an increase of either temperature or density in the local region of the corona. 
Meanwhile, we found that the filter ratio value 
that is a determination factor of temperatures suddenly increases by a factor of about 1.8 when the plasma ejected in the chromosphere above the LB. This indicates that plasma may be temporarily heated up to coronal temperatures. This is consistent with the previous studies of jet-like features observed outside sunspots.
\citet{bharti15} reported an enhanced brightening along the leading edges of jet-like features above a LB with the monochromatic images taken by the SDO/AIA and the IRIS. In addition, \citet{morton12} found an enhanced emission seen along the leading edges of chromospheric jets, which were observed at the edges of the sunspot. \citet{morton12} and \citet{bharti15} suggested that the leading edges of the jets can be heated up to the transition region and coronal temperatures caused by the shock heating. 

In summary, our findings suggest that shock waves can be excited by the local disturbance caused by magnetic reconnection between the emerging flux inside a LB and the adjacent umbral magnetic field. The disturbance generates slow-mode waves, which soon develop into shock waves, and manifest themselves as the arcsecond-scale plasma ejections in the chromosphere. The shock waves we detected also dissipate their energy into heat in the corona. This is consistent with our previous finding that the magnetic reconnection near the photosphere may play an important role in the coronal heating \citep[e.g.][]{song15}. High-resolution coordinated observations between the ground-based telescope, such as the 1.6 meter NST of BBSO, and satellite telescope, such as the IRIS and the SDO/AIA, will be required to clearly understand the dynamic connection from the photosphere to the corona above the LB. 

\acknowledgments

We greatly appreciate the referee's constructive comments. This work was supported by the National Research
Foundation of Korea (NRF-2012 R1A2A1A 03670387). KSC is supported by the "Planetary system research for space exploration" project of the Korea Astronomy and Space Science Institute.



\end{document}